\documentclass[fleqn,11pt]{SelfArx} 

\usepackage{lipsum} 

\usepackage{multirow}

\definecolor{color1}{RGB}{0,0,90} 
\definecolor{color2}{RGB}{0,20,20} 

\usepackage{hyperref} 
\hypersetup{colorlinks=true,urlcolor=blue,linkcolor=blue,linkbordercolor=blue,breaklinks=true,bookmarksopen=false,pdftitle={Title},pdfauthor={Author}}

\JournalInfo{Geomagnetism and Aeronomy, 2022, Vol. 62, No. 7, pp. 919--923.} 
\Archive{DOI: 10.1134/S0016793222070076} 
\Archive{}

\PaperTitle{
Cyclic Variability in Brightness of the Young Solar Analog BE Ceti
} 

\Authors{N.\,I.~Bondar'\textsuperscript{1,*}, and M.\,M.~Katsova\textsuperscript{2,**}
} 

\affiliation{\textsuperscript{1}\textit{
Crimean Astrophysical Observatory, Russian Academy of Sciences, Nauchny, 298409 Russia
}} 
\affiliation{\textsuperscript{2}\textit{
Sternberg State Astronomical Institute, Lomonosov Moscow State University, Moscow, 119991 Russia
}} 
\affiliation{\textsuperscript{*}e-mail: otbn@mail.ru}
\affiliation{\textsuperscript{**}e-mail: maria@sai.msu.ru}
\affiliation{Received February 25, 2022; revised March 27, 2022; accepted May 5, 2022}

\newcommand{\keywordname}{Keywords}

\Abstract{
BE Cet is a young solar analog with an age of 0.6 Gyr and a rotation period of 7.655 days. According
to chromospheric and photospheric indices, its activity is higher than the solar one. An analysis of photometric
data on the time interval between 1977 and 2019 shows the presence of only 6.76 yr cyclic variations in the
mean brightness with an amplitude of 0.02 mag. The obtained cycle is 1--2 yr shorter in comparison with the
chromospheric cycle determined earlier, whose length was estimated to be 9 or 7.6 yr. Parameters of the cycle,
its amplitude and duration change slightly in different epochs. The short-term light variations due to rotational
modulation occur with an increase in amplitude up to 0.05~mag near the activity cycle minimum and
a decrease in its maximum. Some events of a rapid increase in brightness of 0.2--0.6 mag may be considered
as flares.

~

\textbf{DOI} ~ 10.1134/S0016793222070076
}

\begin{document}

\flushbottom 

\maketitle 


\thispagestyle{empty} 

\section{INTRODUCTION}

A study of activity in solar-type stars and its features
in objects with an age younger and older than the
Sun is important for understanding the evolution of
activity and solving some related problems, including
the prediction of the level of activity. One of the first
stars considered in terms of its physical and photometric
parameters as an analog of the Sun is BE Cet (G2.5 V,
$V = 6.39$). This is a bright star in the solar vicinity
located at a distance of 20.4 pc, its age is about 600~Myr
(Hardorp, 1978; Cayrel de Strobel et al., 1981; Eggen,
1960; Cutispoto et al., 2003).

The chromospheric activity of the star has been
studied since 1966 in the framework of the HK-Project
(Baliunas et al., 1995) and is still studied in modern
observational surveys of the Mount Wilson (MW) program
stars. The value of the $\langle S\rangle$-index for BE Cet is
0.35, which is twice higher than that for the Sun. The
S-index varies cyclically. Baliunas et al. (1995) found
the 9-yr main cycle and also suspected the presence of
a long-term trend of 22 yr. Similar results were
obtained by Boro Saikia et al. (2018). Analysis of the
36-yr data series performed by Olah et al. (2016)
allowed them to determine the presence of the main
cycle of 7.6 yr, several short cycles, from 2 to 5 yr, and
a long-term trend.

An increase of the S-index is accompanied by a
decrease in brightness of BE Cet, that is typical for
young stars with a high level of activity. In contrary,
the Sun and stars of comparable age demonstrate a
correlation between the S-index and the photometric
brightness variability (Radick et al., 1998).

A search for the cycles produced by the development
of cool spots on the BE Cet surface has been carried
out only by Messina and Guinan (2002). The
authors constructed a light curve from the photometric
data obtained by some observers in 1986--2000 and
found a low-amplitude cycle of 6.7 yr, they also did
not reveal any other cycles and a long-term trend.

The large series of data on the brightness of the star
accumulated in modern photometric catalogs make it
possible to continue such studies. This work is aimed
to consider the manifestations of activity on the star
from the long-term photometric observations, to identify
activity cycles, rotational modulation effects, and
possible flares. The results of a search for cyclic
changes in the light curve combined from the photometric
data of 1977--2019 are presented in Section 2,
rotational modulation and manifestation of flare
activity are considered in Section 3, Section 4 contains
conclusions.

\begin{figure}[!th] 
\centering
\includegraphics[width=\linewidth]{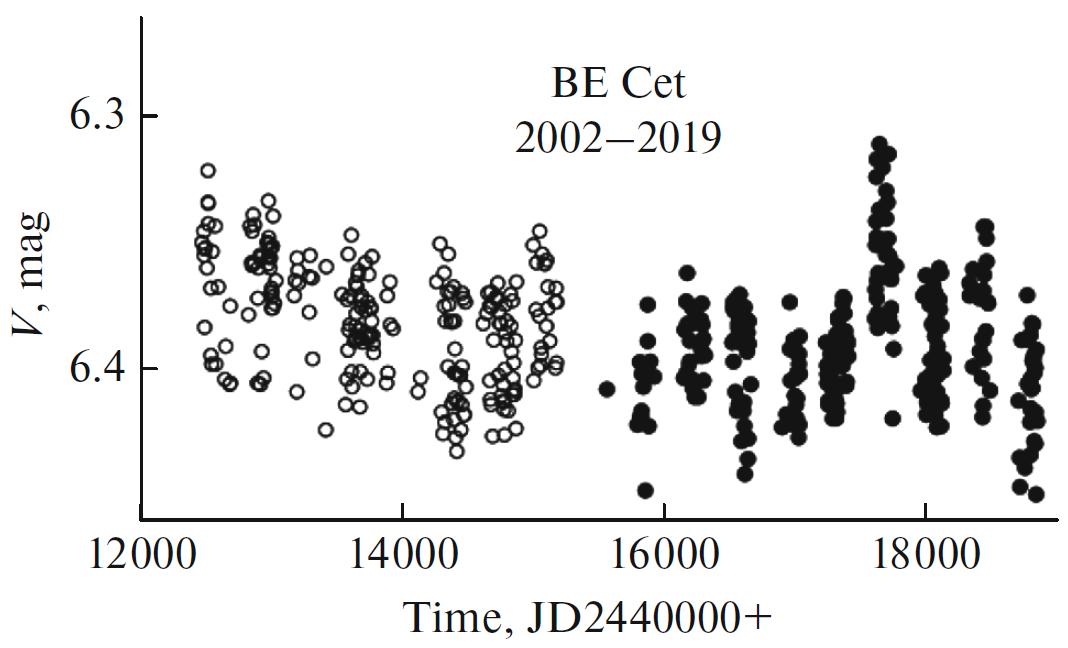}
\vspace*{0.1cm}
\caption{
The light curve of BE Cet on the time interval
2000--2019. $V$-magnitudes from the ASAS catalog (open
circles, N = 230) cover the time span 2002--2009, data for
the time interval 2010--2019 are taken from the KWS survey
(filled circles, N = 276).
}
\label{Figure1}
\end{figure}

\section{CYCLIC VARIABILITY OF PHOTOSPHERIC ACTIVITY}

The photometric series considered by Messina and
Guinan (2002) covers a 14-yr interval. The conclusions
made by the authors about a cycle of 6.7 yr and
the absence of the 22-yr trend differ from the results of
long-term studies of chromospheric cycles of activity
(Baliunas et al., 1995; Boro Saikia et al., 2018; Olah
et al., 2016). To clarify whether these differences are
due to the duration of the photometric series and to
study a stability of the 6.7-yr cycle, we have compiled
a new longer and more complete $V$-magnitude dataset
on the time interval between 1977 and 2019. We have
used the published results and data from the catalogs
of wide-field photometric surveys. Data from the All-Sky 
Automated Survey (ASAS) catalog (Pojmanski,
1997) cover a span of 2002--2009, a sample over 2010--2019 
is taken from the Kamogata Wide-field Survey (KWS) catalog 
[\url{http://kws.cetus-net.org/~maehara/VSdata.py}]. 
The data of both catalogs were
refined, rough estimates and measurements with large
errors were excluded. The resulting light curve over
2002--2019 consists of several data sets that are separated
in time (Fig. 1). Messina and Guinan (2002)
provided the mean $V$-values for 24 observational
epochs in 1986--2000. We have calculated mean
$V$-magnitudes for 29 epochs in 2002--2009 from ASAS
and KWS and 3 epochs from the data obtained by
Chugainov (1980) in 1977--1978. Some information
about the data can be found below in Table 1.

\begin{figure*}[!th] 
\centering
\includegraphics[width=\linewidth]{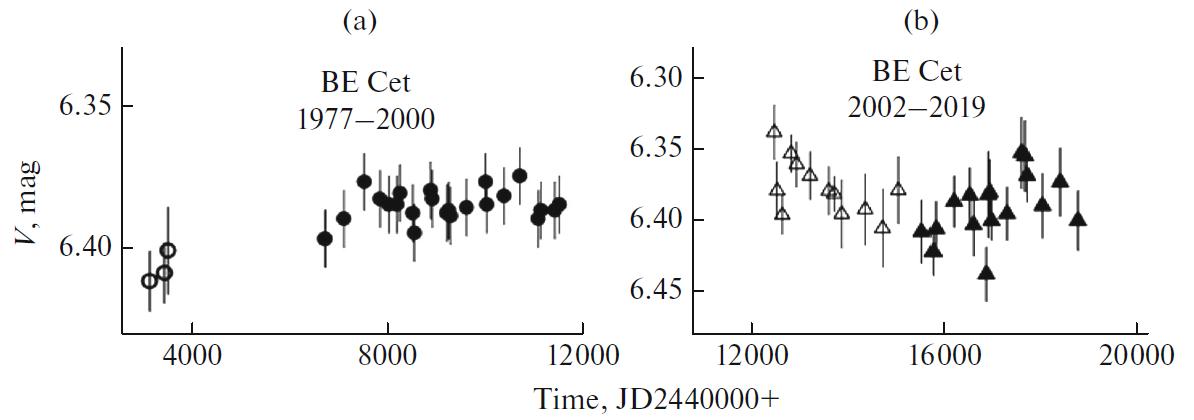}
\includegraphics[width=\linewidth]{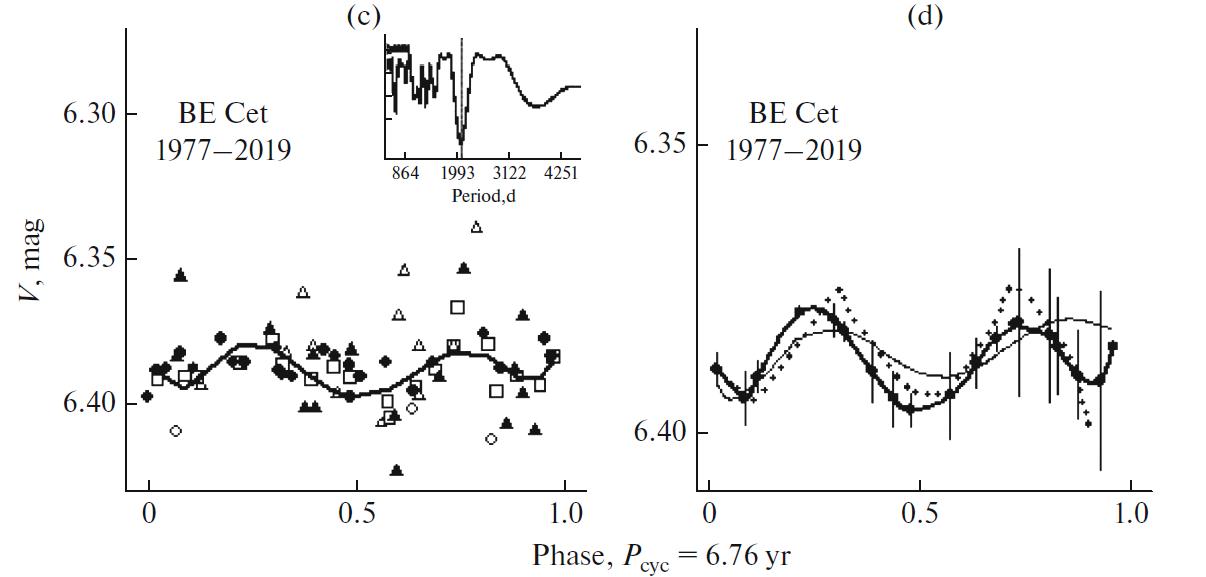}
\vspace*{0.1cm}
\caption{
Seasonal variations in brightness of BE Cet. The light curve in plot "a" represents the results obtained by Chugainov (1980)
in 1977--1978 (open circles) and photometry for 1986--2000 (filled circles) collected by Messina and Guinan (2002). Panel "b"
shows the data from the ASAS catalog (open triangles) and from the KWS catalog (filled triangles). The insert in the lower panel
"c" shows the Scargle periodogram and the presence of $P_{\rm cyc} = 6.76$\;yr. The cyclic variability of the seasonal 
$V$-values with this period are showed in Fig. 2c, symbols are the same as in Figs. 2a, 2b. 
The solid line represents the fitting of the binned $V$-values
(denoted by squares) by a high-degree polynomial. Plot "d" shows the changes in amplitudes and phase shifts for cycles over
1977--2019 (bold line), 1986--2000 (thin line) and 2000--2019 (dotted line) approximated by high-degree polynomials. 
The vertical bars in plots "a, b, d" correspond to the values of $2\sigma$.
}
\label{Figure2}
\end{figure*}

Thus, for further analysis we have obtained 56 values
of the seasonal $V$-magnitudes. The standard deviations
of photoelectric data are less than 0.01 mag, but wide-field
photometry of the stars brighter than 7 mag has
large errors caused by saturation. In our case, the standard
deviations for $V_{\rm season}$ reach 0.02 mag, in some
epochs increase up to 0.023--0.03 mag. Figure 2 shows
changes in brightness of BE Cet from season to season in
the years 1977--2000 (Fig. 2a) and 2002--2019 (Fig. 2b).
Observations in 1977--1978 performed by Chugainov
(1980) are included into consideration to estimate a
range of variations in brightness on the long time span.

The search for the cycle was carried out for all the
data and for the time intervals 1986--2019 and 2002--2019 
by the Hartley and Scargle methods using the
AVE program (Barbera, 1998). The periodogram in
the inset in the Figure 2c shows one significant peak,
which corresponds to $P_{\rm cyc} = 2469.2939$\;days or $\sim 6.76$\;yr
with the error $\sigma = \pm0.55$. Considering the accuracy of
determining the cycle length, its new value does not
differ from $P_{\rm cyc} = 6.7 \pm 0.7$\;yr obtained earlier by Messina
and Guinan (2002). But the phase and amplitude
of the cycle in 1986--2000 some differ from these
parameters for the cycle in 2002--2019. The phases are
computed with the photometric ephemeris ${\rm HJD} =
2446741 + P \times E$. Figure 2c shows convolutions of all
seasonal $V$-magnitudes with the 6.76-yr period and the
approximation of data averaged in bins of 0.05 phase by a
high-degree polynomial. The data of each time series
are indicated by the selected symbols. The cyclic light
curves fitted with polynomials clearly shows the phase
shift and amplitude differences in different epochs
(Fig. 2d). The amplitude of the cycle in 1986--2000 is
about 0.02 mag and it slightly increases up to 0.03 mag
in 2002--2019. Taking into account that the photometric
data over 1977--2000 are more accurate than the
catalog data, the problem of stability of the cycle
parameters remains topical for further consideration.

\begin{figure*}[!th] 
\centering
\includegraphics[width=\linewidth]{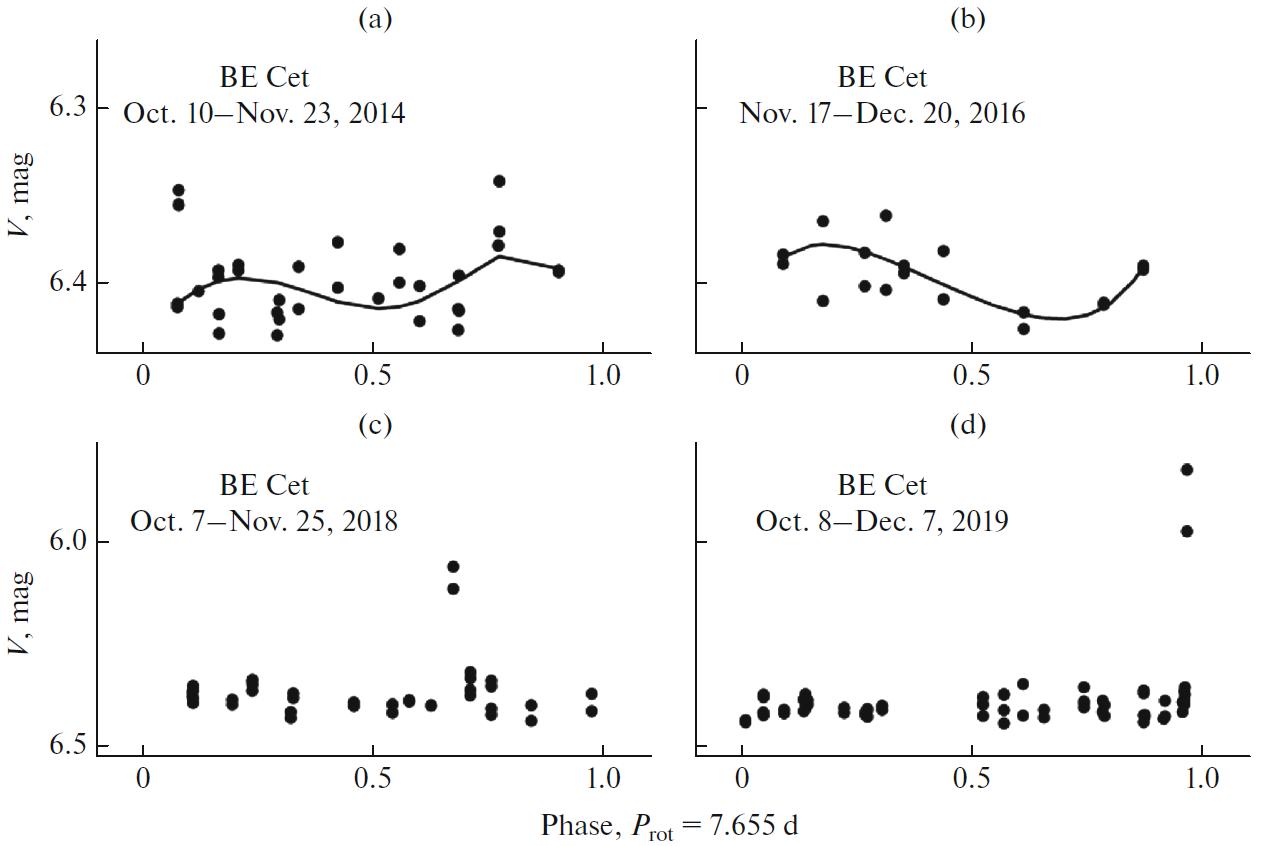}
\vspace*{0.1cm}
\caption{
Rotational modulation and possible flares. The upper panel represents $V$-magnitudes folded with a period of 7.655 days
and polynomial fits of light curves at the indicated intervals chosen near the maximum activity of the 6.76-yr cycle on the time
span JD2456948-2456985 (a) and on the interval of JD2457710-2457742 (b) near to a minimum. Optical flare events are visible
in the light curves on the selected time intervals JD2458399-2458447 (c) and JD2458765-2458824 (d).
}
\label{Figure3}
\end{figure*}

\section{ROTATIONAL MODULATION AND POSSIBLE FLARES}

Low-amplitude periodic variations in brightness of
BE Cet with a period of 7.655 d have first been found
by Chugainov (1980) from his photoelectric observations
in 1977--1978. He explained such changes due to
rotational modulation in brightness produced by cool
spots on the stellar surface. The results published in
the following papers confirm this conclusion. Cutispoto
(1991, 1995), Cutispoto at al. (2003) detected the
rotational modulation in brightness of this star with an
amplitude varying in different epochs from 0.02 to
0.05 mag, sometimes the rotational modulation
amplitude did not exceed the observational errors
(Stepien and Geyer, 1998).

The rotational modulation amplitude is an input
parameter for modeling parameters of starspots and
their distribution on the surface. We consider the rotational
modulation for a series of data from the KWS
catalog on the selected time intervals with denser data
(Figs. 3a, 3b). Phases are computed with a rotational
period of 7.655 days. The light curves are plotted for
the intervals near the activity maximum of the star in
2014 and near the brightness maximum (i.e. the minimum
of the star activity) in 2016. The rotational modulation
amplitude reaches 0.05 mag in epochs near the
maximum of brightness, and it decreases up to 0.03
mag at the minimum, which indicates a more uniform
distribution of starspots in the epoch close to the maximum
activity of the star (Fig. 3b).

Figures 3c, 3d present the series observations for
the epochs when the star suddenly becomes brighter by
a few tenths of magnitudes. The increasing values of
$\Delta V = \langle V\rangle - V_{\rm max}$ 
up to 0.2 mag and 0.6 mag were
detected from observations in October--December in
2018 and 2019. However, patrol observations allow us
only to suppose that these events could be flares, they
were detected by a small number of records. As
shown by the phase curves, possible flares appear
outside the minimum phase, i.e. in the areas distanced
from cool spots.

\begin{table*}[!h] 
\begin{center}
\caption{
Sources of $V$-data to the combined seasonal light curve
}
\par\medskip
\begin{tabular}{|l|c|c|c|c|c|c|}
\hline
\multirow{2}*{Source} & \multirow{2}*{Time span, yr} & 
\multirow{2}{2cm}{Time span JD2440000+ } &
\multirow{2}{2cm}{Number of data ($V_{\rm day}$)} &
\multirow{2}{2cm}{Number of seasons} &
\multicolumn{2}{|c|}{Standard deviation}\\
\cline{6-7}
&&&&&
$\sigma (V_{\rm day})$ 
&
$\sigma (V_{\rm season})$\\
\hline
Chugainov, 1980      & 1977--1978 & 3157--3527   & 50  & 3 & 0.07   & \multicolumn{1}{|c|}{0.01}  \\
Messina et al., 2002 & 1986--2000 & 6741--11552  &     & 24 &       & 0.01  \\
ASAS                 & 2002--2009 & 12478--15067 & 230 & 12 & 0.023 & 0.019 \\
KWS                  & 2010--2019 & 15548--18769 & 276 & 17 & 0.024 & 0.021 \\
\hline
\end{tabular}
\end{center}
\label{Table1}

\end{table*}

\section{CONCLUSIONS}

We have collected all the available photometric
data on the time interval 1977--2019, using the published
data and obtained from the ASAS and KWS
databases to study the photospheric activity of BE Cet
produced by starspots. Analysis of the light curve has
shown the presence of the 6.76 yr cycle and no 
manifestation of long-term variability. The amplitude 
of the cycle is estimated to be 0.02--0.03 mag.

On the entire investigated time interval the peak-to-peak 
seasonal variations of $V$-magnitudes are in the
range of 6.33--6.43 mag, and the yearly mean magnitudes
change in the range of 6.35--6.41 or by 0.07 mag.
The mean $V$-magnitude on the time span 1986--2000 is
6.38 mag, and on the interval 2002--2019 it is 6.39 mag or
remains constant. Our study of photometric behavior
of the star has been carried out on a more extended
time span than in the research by Messina and Guinan
(2002). The obtained results are in good agreement
and give reliable evidence for the absence of a long-term
trend on the considered time span.

The rotational modulation amplitude varies from
season to season and increases in the epochs near the
activity cycle minimum (approaching the brightness
maximum) up to 0.05 mag.

Surface inhomogeneities on G0--G5 stars are rather
difficult to detect by ground-based photometric methods
due of their low contrast relative to the surface
brightness. BE Cet is the first star of an earlier spectral
type than G8, in which BY Dra type variability has
been detected (Chugainov, 1980). As follows from our
result the star also exhibits flare activity. Patrol 
observations have detected events of rapid increases in
brightness by a few tenths of magnitude, which can be
considered as possible optical flares (Bondar' et al.,
2021). We indicated only the events which were confirmed
by two or more records. Small number of
records and short time of observations about 2 min
with a time resolution of 30--35 s are not allow us to
determine a duration of flares.

According to the chromospheric activity indices,
the star BE Cet, an analog of the young Sun at the age
of 600 Myr, belongs to the cool stars with a higher activity
level relative to the modern Sun (Radick et al., 2018). The
long-term photometric study shows that, unlike the
Sun, the duration of the chromospheric activity cycle
of the star is 1--2 yr longer than the 6.76 yr cycle
caused by the development of starspots.

\section{ACKNOWLEDGMENTS}

We have used information from the International Variable
Star Index (VSX) database supported by AAVSO,
Cambridge, Massachusetts, USA, and we are thankful to all
the staff providing the replenishment of these databases and
access to the All-Sky Automated Survey and the 
Kamogata--Kiso--Kyoto Wide-Field Survey. The authors would like to
thank the referees for useful comments and suggestions.

\section{FUNDING}

MK acknowledges the support of Ministry of Science
and Higher Education of the Russian Federation, grant
no. 075-15-2020-780.

\section{CONFLICT OF INTEREST}

The authors declare that they have no conflicts of interest.

\bibliographystyle{unsrt}

\end{document}